# PECULIARITIES OF THE CORRELATION BETWEEN LOCAL AND GLOBAL NEWS POPULARITY OF ELECTRONIC MASS MEDIA

Lande D.V. [1, 2], Braichevskii S.M. [1], Darmokhval A.T. [1], Snarskii A.A. [2]

[1] – Elvisti Information Center, Kyiv, Ukraine,
[2] - Keiv Polytechnical Institute, Kyiv, Ukraine

*One of the approaches to the solution of the navigation problem in current information flows is ranking the documents according to their popularity level. The definition of local and global news popularity which is based on the number of similar-in-content documents, published within local and global time interval, was suggested. Mutual behavior of the documents of local and global popularity levels was studied. The algorithm of detecting the documents which received great popularity before new topics appeared was suggested.*

Information flows, generated into the Internet, condition the problem of navigation in network resources [1]. One of the approaches to the solution of this problem is ranking the documents according to their popularity level. On the other hand, information flows are mostly caused by the events taken place in a real world. In fact, one can imagine the appearance of information flows as the generation and movement of data sets associated with a definite event, being realized as a sense part. Indefinite number of messages may correspond to one event. So, characteristics of information flows are primarily determined by event flows of a real world. That is why it is appropriate to speak about the correlation which exists between news popularity and event poularity, switching from the analysis of news flows to that of events.

We may speak about the popularity of the news taking into account the number of messages similar to the one under consideration. There are many definitions of formal similarity which are used in retrieval systems in the regimes "search of similar documents". In particular, the authors used the method of defining the similarity, applied in InfoStream system [2]. The message is considered to be similar to the initial one if it contains a certain number of the most significant words from it (let us call this criterion $\alpha$-similarity). The principle of detecting significant words and their number is based on both statistical algorithm, built on Tsypf regularity, and on some empirical-linguistical approaches.

Global popularity for each news ($P_G$) refers to the number of $\alpha$-similar news in a retrospective database (over ten million documents for the last year). Local popularity ($P_L$) refers to the number of $\alpha$-similar news on the day when the initial news appeared.

To illustrate the correlation between local and global popularity, the authors considered a thematic file of the documents during three days; it contained 5000 messages/news and was ranked according to global popularity $P_G$. The dependence of $P_G$ popularity on the numbers of the documents is presented in Fig.1. Each number of the documents, in a ranked $P_G$ sequence, has some meaning of $P_L$ popularity (Fig. 2). The behavior of these dependences appears to differ considerably.

However, the figures presented here show that some correlations exist between dependences under consideration. The correlation dependences by their properties appeared to be close to fractal

[3, 4]. In particular, the nature of sophisticating local popularity when global popularity rises (Fig. 3) resembles visually the realization of the algorithm of developing a fractal structure.

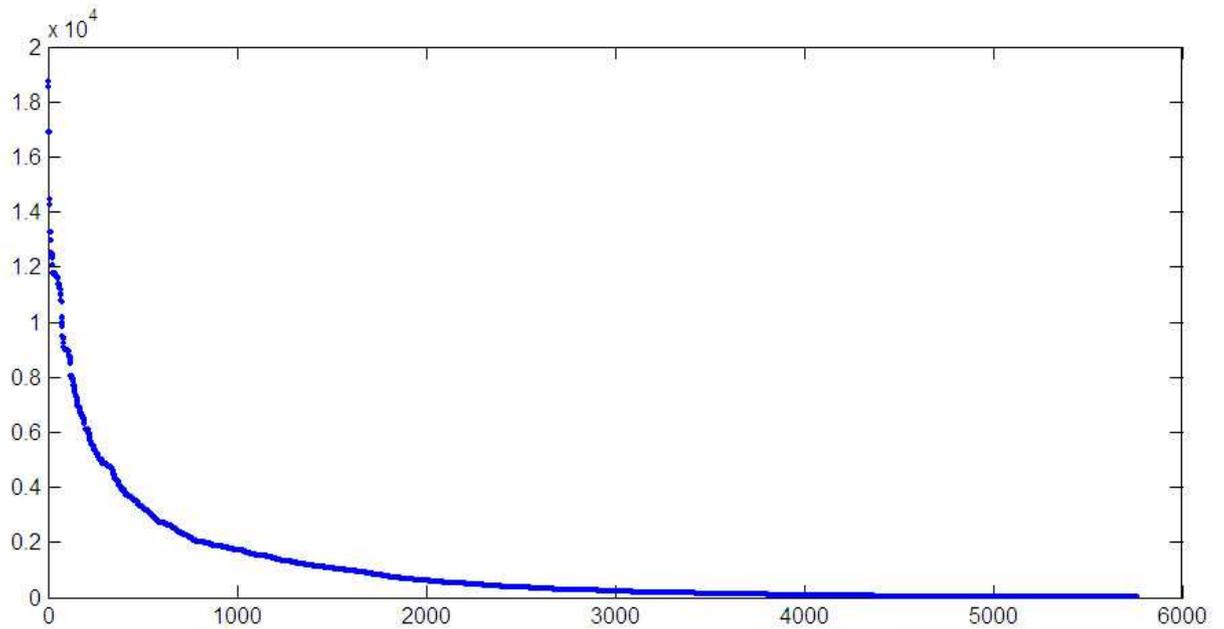

*Рис. 1. Meanings of popularity $P_G$ (Y-axis) as to the numbers ranked*

*on $P_G$ documents (X-axis),*

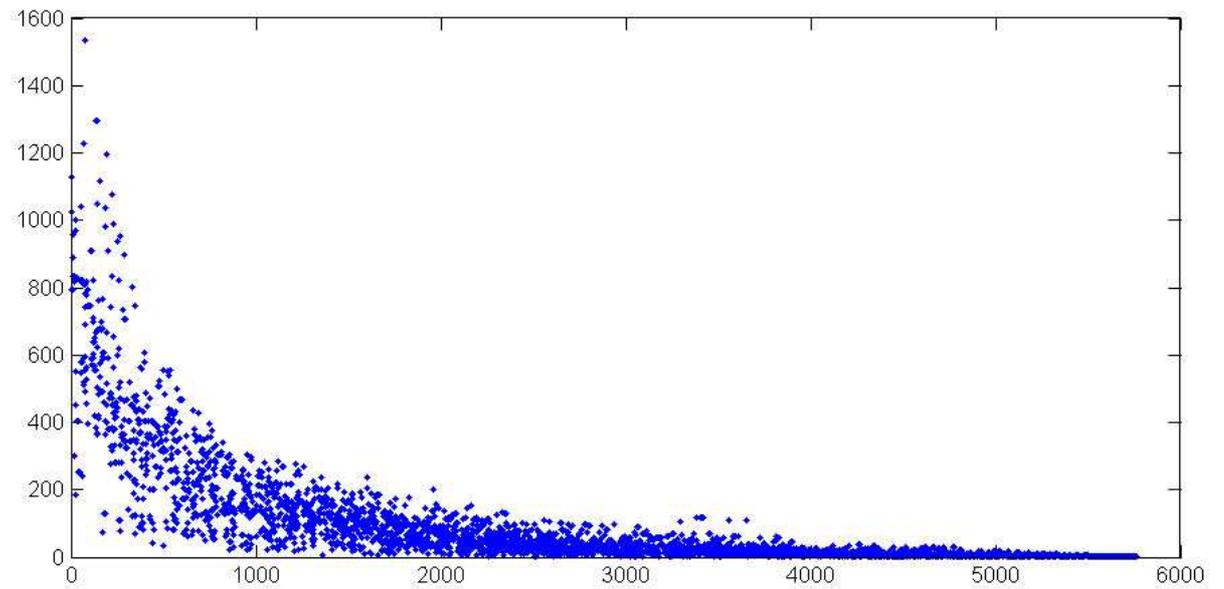

*Fig. 2. Meanings of popularity $P_L$ (Y-axis) as to the numbers ranked*

*on $P_G$ documents (X-axis),*

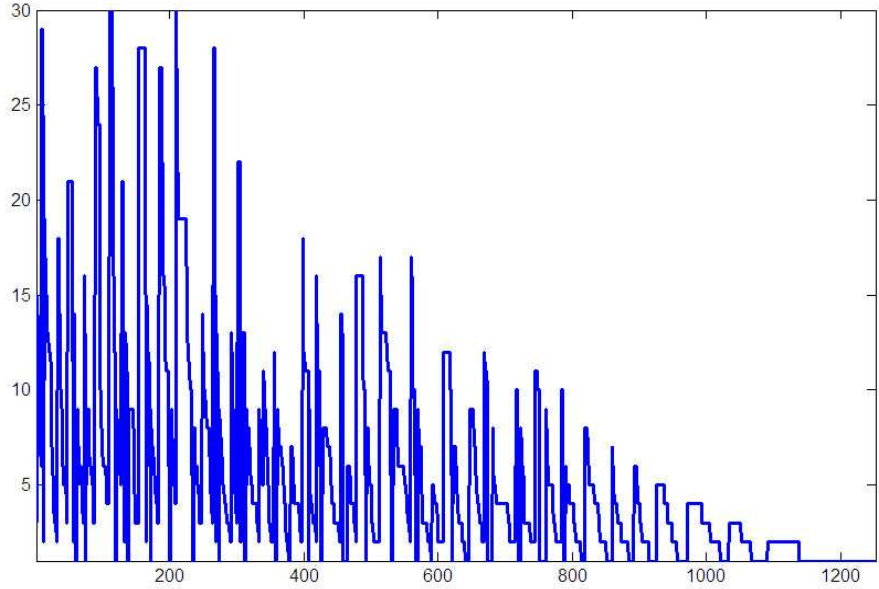

*Fig. 3. Sophisticating the structure of local popularity levels when global popularity rises*

Fig. 4 shows the correlation between local popularity and global popularity ($K_{new} = P_L/P_G$), for some news this indicator is rather high. This fact allows to classify the events described in the news as the new ones.

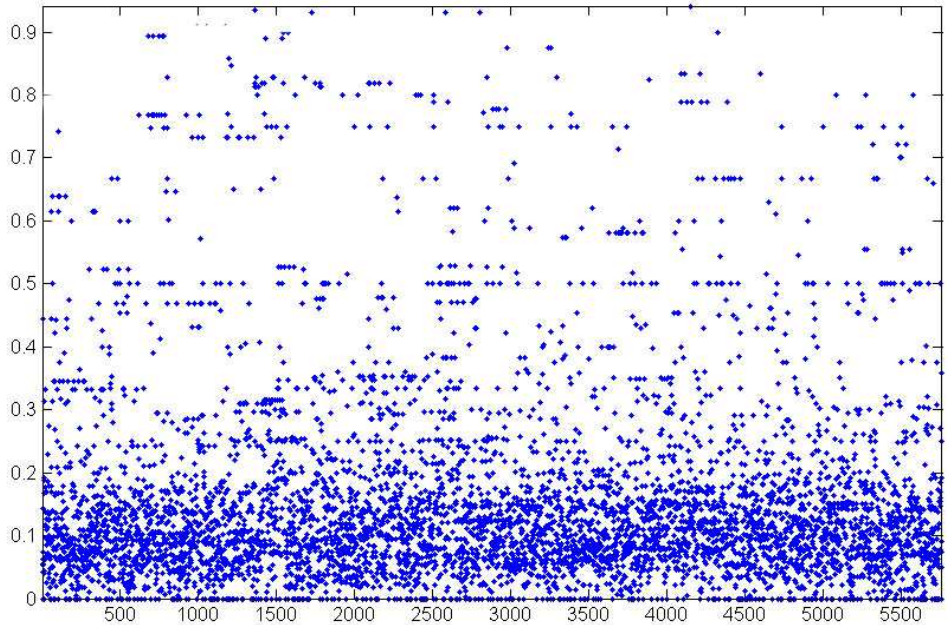

*Fig. 4. Sophisticating the structure of local popularity level when global popularity rises*

Fractal properties of the consequence of the topic novelty of the news ($K_{new}$) are conditioned by the nature of generating new messages in electronic mass media. There exist several approaches to defining fractal characteristics of measurement levels, one of them being DFA (Detrended Fluctuation Analysis) [4]; it is a type of the analysis of variance when root-mean-square error of linear approximation depending on the size of approximation segment is analyzed. In the framework of this method, first of all data are given (in the case under consideration, meanings of $K_{new}$ to zero mean, as a result of which consequence $y(k)$ is built). Then the sequence is broken into continuous length segments $n$, within each of them the equation of line, approximazing the sequence $y(k)$, is defined using the method of least squares.

The approximation found $y_n(k)$ ($y_n(k) = ak + b$) is considered to be a local trend. Then root-mean-square error of linear approximation, when meaning range $n$ is wide, is calculated. In case, when dependence $D(n)$ has power nature $D(n) \sim n^\alpha$, i.e., the availability of linear segment with double logarithmic scale $\ln D \sim \alpha \ln n$, it is appropriate to mention the existence of scaling.

As Fig.5 shows, dependence $D(n)$ for sequence $K_{new}$ depends on $n$ in power way, i.e., in double logarithmic scale this dependence is closer to a linear one. This fact confirms that in real documentary files steady non-trivial relationship between local and global popularity of documents exists.

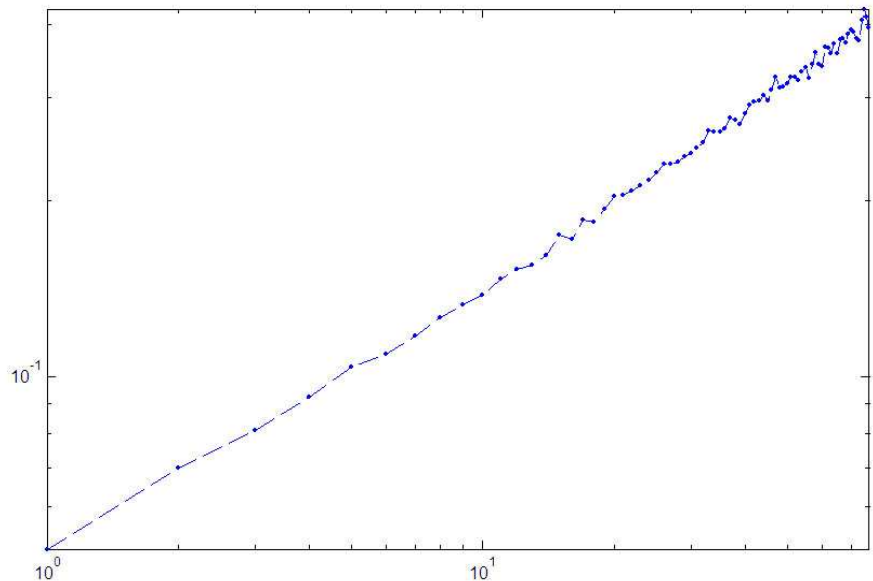

*Fig. 5. Dependence D(n) of observation row (Y-axis) on the length of approximation segment*

*(X-axis) in a logarithmic scale*

Hence, a method of detecting news which got higher popularity recently was worked out (i.e., the news which meanings $K_{new}$ exceed the threshold, planned by experts), which is partial solution of urgent problem of new event detection that has been actively discussed by world specialists for several years [5, 6]. The practice of introducing the mechanism of new event detection in the framework of InfoStream technology confirms its efficiency as an essential contribution to retrieval regimes. Apparently, the most important point here is that the user feels more attached to new events of a real world rather than to new documents, in other words, "a semantic shift" of perception occurs.

**Reference**
1. Braichevskii, S.M. Lande, D.V. Urgent aspects of current information flow // Scientific and technical information processing / - Allerton press, inc. 2005, Vol 32; part 6, pp. 18-31.
2. Lande, D., Darmokhval A., Morozov A. The approach to duplication detection in news information flows // In Proc. Of the 8th Russian Conference on Digital Libraries

RCDL'2006, Suzdal, Russia, 2006. – P. 115-119. Available online: http://www.rcdl2006.uniyar.ac.ru/papers/paper_71_v2.pdf
3. Feder J. Fractals. New York: Plenum Press, 1988.
4. Peng C.K., Havlin S., Stanley H.E., Goldberger A.L. Quantification of scaling exponents and crossover phenomena in nonstationary heartbeat time series, Chaos., 5, 82, 1995.
5. Papka, R. On-line News Event Detection, Clustering, and Tracking. Ph. D. Thesis, University of Massachusetts at Amherst, September 1999.
6. Lande, D.V. Furashev, V.N. New Event Detection in the Framework of a Content Monitoring System // Scientific and technical information processing / - Allerton press, inc. 2006, Vol 40; Part 6, pp. 239-243.